\documentclass[10pt]{article}
\usepackage[dvips]{graphicx}
\usepackage[tight]{subfigure}

\def\be{\begin{equation}}
\def\ee{\end{equation}}
\def\bea{\begin{eqnarray}}
\def\eea{\end{eqnarray}}
\def\beaN{\begin{eqnarray*}}
\def\eeaN{\end{eqnarray*}}
\def\ed{\end{document}}
\def\bit{\begin{itemize}}
\def\eit{\end{itemize}}

\def\sig{\sigma}

\def\lam{\lambda}

\def\Del{\Delta}

\def\k{\kappa}
\def\alf{\alpha}

\def\BD{\Bar D}

\def\di{\partial}

\def\half{{\textstyle{1 \over 2}}}
\def\~{\tilde}
\def\lag{{\hat{\cal L}}}
\def\m{\label}
\def\l{\left}
\def\r{\right}

\def\goto{\rightarrow}
\def\Bar{\overline}
\def\const{\rm const}

\begin{document}

\centerline{\bf THREE TYPES OF SUPERPOTENTIALS FOR} \centerline{\bf
PERTURBATIONS IN THE EINSTEIN-GAUSS-BONNET GRAVITY}

\smallskip

\centerline{\it A.N.Petrov}

\smallskip

\centerline{\it Relativistic Astrophysics group, Sternberg
Astronomical institute,} \centerline{\it
 Universitetskii pr., 13, Moscow, 119992,
RUSSIA }

\smallskip

\centerline{ e-mail: anpetrov@rol.ru}

\smallskip

telephone number: +7(495)7315222

\smallskip

fax: +7(495)9328841

\smallskip

PACS numbers: 04.50.+h, 04.20.Cv, 04.20.Fy

\begin{abstract}
Superpotentials (antisymmetric tensor densities) in
Einstein-Gauss-Bonnet (EGB) gravity for arbitrary types of
perturbations on arbitrary curved backgrounds are constructed. As a
basis, the generalized conservation laws in the framework of an
arbitrary D-dimensional metric theory, where conserved currents are
expressed through divergences of superpotentials, are used. Such a
derivation is exact (perturbations are not infinitesimal) and is
approached, when a one solution (dynamical) is considered as a
perturbed system with respect to another solution (background).
Three known prescriptions are elaborated: these are the canonical
N{\oe}ther theorem, the Belinfante symmetrization rule and the
field-theoretical derivation. All the three approaches are presented
in an unique way convenient for comparisons and a development. Exact
expressions for the 01-component of the three types of the
superpotentials are derived in the case, when an arbitrary static
Schwarzschild-like solution in the EGB gravity is considered as a
perturbed system with respect to a background of the same type.
These formulae are used for calculating the mass of the
Schwarzschild-anti-de Sitter black hole in the EGB gravity. As a
background both the anti-de Sitter spacetime in arbitrary dimensions
and a not maximally symmetric ``mass gap'' vacuum in 5 dimensions
are considered. Problems and perspectives for a future development,
including the Lovelock gravity, are discussed.
\end{abstract}

\eject

\section{Introduction}
\m{Introduction}

Multidimensional (multi-D) metric theories of gravity become very
popular (see, for example, reviews \cite{Rubakov} - \cite{Obers} and
references there in). It is important to construct conserved
quantities for perturbations in such theories. Several useful and
interesting approaches were presented (see, for example,
\cite{Paddila} - \cite{Olea1}). However, because these theories and
their solutions are very complicated and have nontrivial properties,
one needs to expand a research of conservation laws. In particular,
curved backgrounds (including not symmetrical, {\em etc.}) are
usual, therefore a construction of covariant conserved quantities on
such backgrounds is necessary. Already, in 4-dimensional general
relativity (4D GR), there exist approaches satisfying this
requirement. From them three ones are more universal and satisfy
necessary tests in 4D GR, when one calculates masses of black holes,
energy and momentum fluxes in the Bondi-Sachs solution, {\em etc.}.
It is natural to develop these three approaches in multi-D theories.

One of the approaches, {\em canonical}, begins from the Einstein
pseudotensor \cite{Einstein16} and the Freud superpotential
\cite{Freud39}. Its generalized form in 4D GR is presented by  Katz,
Bi\v c\'ak  and Lynden-Bell (KBL) \cite{KBL}. The second approach is
based on the {\em Belinfante symmetrization} method
\cite{Belinfante}, which firstly in 4D GR has been applied by
Papapetrou \cite{Papapetrou48} for symmetrization of the Einstein
pseudotensor. A generalized application of the Belinfante method in
4D GR is presented in \cite{PK}. At last, the third approach starts
from the linearization of the Einstein equations and treating all
the other terms as an effective energy-momentum
\cite{Weinberg-book}. This picture has been recreated in the form of
the Lagrangian based field theory for perturbations on a flat
background by Deser \cite{[11]}, so-called {\em field-theoretical}
derivation, which later has been developed for arbitrary curved
backgrounds \cite{AbbottDeser82} - \cite{Petrov2008}. All the three
methods derive the exact (not infinitesimal or approximate)
perturbations. Such a derivation is approached when a one solution
(dynamical) of the theory is considered as a perturbed system with
respect to another solution (background) of the same theory. In the
present paper we follow just this scheme and present exact formulae.
Linear or of higher order approximations easily follow if the exact
form is presented.

In multi-D theories, the above mentioned approaches already have
received a development as follows. In the works \cite{DerKatzOgushi}
- \cite{DerMor05} Deruelle and Katz with coauthors successfully
develop the canonical KBL approach \cite{KBL}. The method based on
the Belinfante procedure is continued in our works
\cite{PK2003a,Petrov2008}. The field-theoretical derivation  in
multi-D arbitrary gravities  starts in \cite{[15]}. In the recent
works \cite{DT2} - \cite{DT3} Deser and Tekin, applying the
Abbott-Deser procedure \cite{AbbottDeser82}, suggest a construction
of conserved charges for perturbations about vacua in quadratic (in
curvature) theories. A development of our \cite{[15]} and the
Deser-Tekin \cite{DT2} methods is continued in \cite{Petrov2005b}.
The results in \cite{DerKatzOgushi} - \cite{Petrov2005b} show that
all the three approaches are perspective in constructing conserved
charges for numerous known and new solutions in multi-D theories.

Among multi-D theories, the Gauss-Bonnet modification (quadratic
order of the Lovelock type \cite{Lovelock}) of the Einstein theory
takes a very important place. The main goal of the present paper is
to construct superpotentials of the most general form in the
framework of this theory with using the aforementioned approaches.
As a basis, we use the results obtained in \cite{Petrov2008} in the
framework of an arbitrary multi-D metric theory. Following them, in
the next section \ref{D-dimensions}, we derive out the explicit
generalized expressions for the superpotentials. In section
\ref{EGBsuperpotentials}, we construct superpotentials in the
Einstein-Gauss-Bonnet (EGB) gravity by all the three methods for
arbitrary types of perturbations on arbitrary curved backgrounds.
Thus many new solutions (see, e.g., \cite{Aliev} -
\cite{DottiOlivaT1}) can be examined. Concerning the canonical
method, we in many lines repeat the work \cite{DerKatzOgushi},
however, we present it in the unique style of the present paper. In
section \ref{Ssss}, the developed formalism is used for an
examination of static spherically symmetric solutions of the
Schwarzschild-like type in EGB gravity. In spite of a visible
simplicity,  such solutions are very rich in interesting properties.
For calculating a mass we derive out {\em exact} expressions for the
$01$-component of the superpotentials of all the three types,
considering background solutions of the same Schwarzschild-like
form. In section \ref{BHinEGB}, for each of the approaches the mass
of the Schwarzschild-anti-de Sitter (S-AdS) black hole (BH)
\cite{BD+} is calculated. In the first case, as a usual background
we consider the AdS spacetime in arbitrary D dimensions. The second
case is not trivial, in 5 dimensions as a not maximally symmetric
background we consider the vacuum of a ``mass gap'' (see, e.g.,
\cite{Cai}). All the results are acceptable. In section
\ref{Discussion}, problems of the methods and perspectives for a
future development are discussed.

\section{Preliminaries}
 \m{D-dimensions}
 \setcounter{equation}{0}

In this section, we shortly represent the results of the section in
\cite{Petrov2008} related to constructing superpotentials in an
arbitrary multi-D metric theory. Consider the Lagrangian:
 \be
 \lag_D = - \frac{1}{2\k}\lag_{g}(g^a) + \lag_{m}(g^a,\Phi)
 \m{lag-g}
 \ee
including derivatives up to the second order of $g^a$ and $\Phi$. An
independent metric variable $g^a$ is thought as a one defined in the
set
  \be
 g^{a} = \l\{g^{\mu \nu },~g_{\mu
\nu },~\sqrt{-g}g^{\mu \nu },~\sqrt{-g}g_{\mu \nu },~(-g)g^{\mu \nu
},~\ldots\r\}\,, \m{(1)}
 \ee
where the Greek indexes numerate $D$-dimensional spacetime
coordinates. Variation of (\ref{lag-g}) with respect to each $g^a$
leads to the gravitational equations, which are equivalent between
themselves. Below we will explain why it is important to examine the
different definitions (\ref{(1)}). Matter sources without
concretization are denoted as $\Phi$; here and below ``hat'' means
densities of the wight +1; $({,\alf}) \equiv \di_{\alf}$ means
ordinary derivatives.

Our goal is to describe perturbations on a background of a fixed
system.  For the last we assume the background $D$-dimensional
spacetime with the metric $\Bar g_{\mu\nu}$, on the basis of which
the background Christoffel symbols $\Bar \Gamma^\sig_{\tau\rho}$,
covariant derivatives $\Bar D_\alf$ and the background Riemannian
tensor $\Bar R^\sig{}_{\tau\rho\pi}$ are constructed; here and below
``bar'' means that a quantity is a background one. We use also the
background Lagrangian defined as $\Bar \lag_D = \lag_D(\bar
g^a,\Bar\Phi)$ and corresponding background gravitational equations.
We set that the background fields $\bar g^a$ and $\Bar\Phi$ satisfy
the background equations, and, thus are known (fixed).

Using different combinations of the pure gravitational Lagrangians
$-\lag_g/2\k$ and $-\Bar\lag_g/2\k$ and applying the N{\oe}ther
procedure directly or in a combination with other methods one
obtains the correspondent identities of the unique form:
  \be
 \hat {\cal I}^\alf(\xi) \equiv \di_\beta \hat {\cal I}^{\alf\beta} (\xi)\,
 \m{generalCLs}
  \ee
where $\xi^\alf$ is a displacement vector, $\hat {\cal
I}^{\alf\beta}$ is an antisymmetric tensor density, thus $\Bar
D_{\alf\beta} \hat {\cal I}^{\alf\beta} (\xi)\equiv \di_{\alf\beta}
\hat {\cal I}^{\alf\beta} (\xi)\equiv 0$; and $\hat {\cal I}^\alf$
is a vector density. After substitution of the dynamical and
background equations into $\hat {\cal I}^\alf$ it acquires the sense
of the conserved current $\Bar D_{\alf} \hat {\cal I}^{\alf} (\xi) =
\di_{\alf} \hat {\cal I}^{\alf} (\xi)= 0$, and the identity
(\ref{generalCLs}) acquires the sense of the conservation law
(already not identity) with the superpotential $\hat {\cal
I}^{\alf\beta}$. Integrating the aforementioned differential
conservation laws and using the D-dimensional Gauss theorem, one can
obtain the integral charges in a generalized form:
 \be
{\cal P}(\xi) = \int_\Sigma d^{D-1} x\,\hat {\cal I}^0(\xi) =
\oint_{\di\Sigma} dS_i \,\hat {\cal I}^{0i}(\xi)\,
 \m{charges}
 \ee
where $\Sigma$ is a spatial $(D-1)$-dimensional hypersurface $x^0 =
\const$, $\di\Sigma$ is its  $(D-2)$-dimensional boundary, zero's
and small Latin indexes numerate time and space coordinates,
respectively. In depending on boundary conditions at $\di\Sigma$ (or
on a falloff of potentials, if $\di\Sigma$ presents infinity), the
integral (\ref{charges}) defines the {\em conserved} on $\Sigma$
quantity $\di_0 {\cal P}(\xi)= 0$ or expresses its {\em flux}
through $\di\Sigma$: $\di_0 {\cal P}(\xi)=-\oint_{\di\Sigma} dS_i
\,\hat {\cal I}^{i}(\xi)$.

As a result of an application of the N{\oe}ther theorem the
quantities in (\ref{generalCLs}) and (\ref{charges}) in general
contain {\em arbitrary} displacement vectors $\xi^\alf$. However, an
interpretation such quantities, as a rule, is impossible or not
understandable\footnote{ Nevetherless, sometimes one can find out a
reasonable using even non-Killing vectors, like in \cite{PK}, where
{\em conformal Killing} vectors were used for derivation of
perturbations on the Friedmann-Robertson-Walker background.},
whereas using the {\em Killing} vectors lead to the clear treating.
Here, to define the energy/mass of the system in our applications we
use the timelike Killing vector of the background.

\subsection{The canonical superpotentials}
\m{canonical}

To construct the canonical expressions we follow the KBL ideology
\cite{KBL}   and consider the Lagrangian:
 \be
 \lag_{G} = -\frac{1}{2\k}\l(\lag_{g} - \Bar\lag_{g} + \di_\alf \hat
 d^\alf\r)\,,
 \m{ArbitraryLagPert}
 \ee
which, of course, leads to the usual gravitational equations. Then
the background metric $\Bar g_{\mu\nu}$ is incorporated into
$\lag_g$ by the usual way as follows. The ordinary derivatives
$\di_{\alf}$ are rewritten over the covariant $\BD_\alf$ ones by
changing $\di_\tau g^a \equiv \Bar D_\tau g^a - \Bar
\Gamma^\sig_{\tau\rho} \l. g^a \r|_\sig^\rho$ where $\l. g^a
\r|^\alf_\beta$ is defined by the transformation properties of
$g^a$. Then, with using
 \bea
\Del^\alpha_{\mu\nu} & = &\Gamma^\alpha_{\mu\nu} - \Bar
{\Gamma}^\alpha_{\mu\nu} = \half g^{\alf\rho}\l( \BD_\mu g_{\rho\nu}
+ \BD_\nu g_{\rho\mu} - \BD_\rho g_{\mu\nu}\r)\, , \m{DeltaDefD}\\
  R^\lam{}_{\tau\rho\sig} & =&
\BD_\rho \Delta^\lam_{\tau\sig} -  \BD_\sig\Delta^\lam_{\tau\rho} +
 \Delta^\lam_{\rho\eta} \Delta^\eta_{\tau\sig} -
 \Delta^\lam_{\eta\sig} \Delta^\eta_{\tau\rho}
 + \Bar R^\lam{}_{\tau\rho\sig}\,
 \m{(17-DmD)}
 \eea
we transform the pure metric Lagrangian $\lag_g$ into an explicitly
covariant form:
 $\lag_g = \lag_c = {\lag}_c (g^a; \Bar D_\alf
g^a; \Bar D_\beta \Bar D_\alf  g^a)$. Then, a direct application of
the {\em canonical} N{\oe}ther procedure to $-\lag_c/2\k $ gives the
identity of the type (\ref{generalCLs}) with the superpotential:
 \be
 \hat {\imath}^{\alf\beta}_C(\xi) =
 \l({\textstyle{2\over 3}}
 \BD_\lam  \hat n_{\sig}{}^{[\alf\beta]\lam}  - \hat
 m_{\sig}{}^{[\alf\beta]}\r)\xi^\sig   -
 {\textstyle{4\over 3}} \hat n_{\sig}{}^{[\alf\beta]\lam}
 \BD_\lam  \xi^\sig \,
 \m{+IsupD+}
 \ee
where the coefficients  are calculated with the use of the general
formulae:
 \bea
  \hat m_\sig{}^{\alf\tau} & \equiv & -\frac{1}{2\k}\l\{
 \l[{{\di \lag_c} \over {\di (\Bar D_\alf g^a)}} -
 \Bar D_\beta \l({{\di \lag_c} \over {\di (\Bar D_\beta \Bar D_\alf g^a)}}\r)\r]
 \l.g^a\r|^\tau_\sig \r.\nonumber \\ & - & \l.
 {{\di \lag_c} \over {\di (\Bar D_\tau \Bar D_\alf g^a)}}
\Bar D_\sig g^a +
 {{\di \lag_c} \over {\di (\Bar D_\beta \Bar D_\alf g^a)}}
 \Bar D_\beta (\l.g^a\r|^\tau_\sig)\r\}\, ,
\m{(+4+)}
 \eea
\be \hat n_\sig{}^{\alf\tau\beta} \equiv -\frac{1}{4\k} \l[{{\di
\lag_c} \over {\di (\Bar D_\beta \Bar D_\alf g^a)}}
 \l.g^a\r|^\tau_\sig +
 {{\di \lag_c} \over {\di (\Bar D_\tau \Bar D_\alf g^a)}}
 \l.g^a\r|^\beta_\sig\r].
\m{(+5+)}
 \ee
To calculate the full superpotential related to the Lagrangian
(\ref{ArbitraryLagPert})  one has to apply the barred procedure to
(\ref{+IsupD+}) and the N{\oe}ther procedure to the divergence in
(\ref{ArbitraryLagPert}):
 \be
 \hat {\cal I}^{\alf\beta}_C(\xi) = \hat {\imath}^{\alf\beta}_C(\xi)
 - \Bar{\hat {\imath}^{\alf\beta}_C(\xi)} +
\k^{-1}\xi^{[\alf}\hat d^{\beta]}\, .
 \m{IsupD}
 \ee

As is seen, if one adds different divergences to the Lagrangian
(\ref{ArbitraryLagPert}), different canonical superpotentials
(\ref{IsupD}) appear. Such a situation is natural and could be
interesting in gravitational theory, when analogies, for example,
with thermodynamics are carried out \cite{ChangNesterChen}. There is
no a crucial principle for a definition of divergences, however the
Silva method \cite{Silva99} looks as a perspective criterium. The
boundary conditions for deriving superpotentials in \cite{Silva99}
could be imposed by including a correspondent divergence into the
Lagrangian. This divergence is defined by an {\em unique} way. Thus,
supposing the boundary conditions for an isolated system in 4D GR
the Silva method shows that the KBL divergence \cite{KBL} is the
{\em unique} answer for the Dirichlet boundary conditions.

The superpotential (\ref{IsupD}) with the KBL divergence in the
Lagrangian derived for the 4D Einstein theory, simplified to the
Minkowski background in the Cartesian coordinates and with the
translation Killing vectors $\xi^\alf \goto \lam^\alf_{(\beta)} =
\delta^\alf_{\beta}$ is just the very known Freud superpotential
\cite{Freud39}. The current corresponding to (\ref{IsupD}) under
these simplifications transforms into the famous Einstein
pseudotensor \cite{Einstein16}.

In a recent work \cite{KatzLivshits}, Katz and Livshits develop
Silva's method applying it in the Lovelock theory \cite{Lovelock} in
the first order formalism (Palatini presentation). Using boundary
conditions correspondent to an isolated system, they derive out
equations for superpotentials, which are easily integrated giving
the general expressions for all the superpotentials associated with
the Lovelock Lagrangians. In particular, in multi-D GR, the
Katz-Livshits superpotential turns out {\em uniquely} the KBL
superpotential; in EGB gravity, they present explicitly a new
superpotential. The last naturally transfers into the KBL
superpotential for $D=4$. Below we are interested in a divergence
induced by the Katz-Livshits procedure in the EGB Lagrangian.
Together with this we consider the divergence presented by Deruelle,
Katz and Ogushi \cite{DerKatzOgushi}, which has been derived ``in
such a way that the boundary term [after variation] does not contain
terms proportional to normal derivatives of the metric variations''.

\subsection{The Belinfante symmetrization procedure}
\m{Belinfante}

The application of the generalized Belinfante procedure
\cite{PK,Petrov2008} to the canonical conserved laws with the
superpotential (\ref{+IsupD+}) leads to the superpotential:
  \be
 \hat {\imath}_{B}^{\alf\beta}(\xi) =
 2
\l({\textstyle{1\over 3}}\BD_\rho \hat n_{\sig}{}^{[\alf\beta]\rho}+
\BD_\tau \hat
 n_{\lam}{}^{\tau\rho[\alf} \bar g^{\beta]\lam} \bar g_{\rho\sig}\r)
\xi^\sig - {\textstyle{4\over 3}} \hat
 n_{\sig}{}^{[\alf\beta]\lam}\BD_{\lam}\xi^\sig\, .
 \m{+(supIB+}
 \ee
Fully the superpotential correspondent to the Lagrangian
(\ref{ArbitraryLagPert}) has the form:
 \be
 \hat {\cal I}_{B}^{\alf\beta}(\xi) =  \hat
 {\imath}_{B}^{\alf\beta}(\xi) -\Bar{ \hat {\imath}_{B}^{\alf\beta}(\xi)}
\, .
 \m{(supIB+}
 \ee
Because this superpotential depends on the $n$-coefficients only it
vanishes for Lagrangians with only the first order derivatives (see
(\ref{(+5+)})) and is well adapted to theories with second
derivatives in Lagrangians, like quadratic \cite{DT2} or of the
Lovelock type \cite{Lovelock} theories.

The superpotential (\ref{(supIB+}) and the correspondent current
derived for the 4D GR, simplified to the Minkowski background in the
Cartesian coordinates and with the translation Killing vectors are
the well known famous Papapetrou superpotential and its current
\cite{Papapetrou48}.

\subsection{The field-theoretical prescription}
\m{f-t}

Turning to the field-theoretical approach one has to use the
decomposition of the metric and matter variables
 \be
 g^a = \Bar g^a + h^a\, ,\qquad
 \Phi= \Bar \Phi + \phi\,.
 \m{g-Dec}
 \ee
Following \cite{[15]} we describe the perturbed system by the
Lagrangian:
 \be
\lag^{dyn}_D = \lag_D (\Bar g+h,\,\Bar \Phi+\phi ) - h^a
\frac{\delta \Bar \lag_D}{\delta \Bar g^a} - \phi\frac{\delta \Bar
\lag_D}{\delta \Bar \Phi}- \Bar \lag_D + div\,.
 \m{lag}
 \ee
Subscript ``${dyn}$'' is used because the perturbations $h^a$ and
$\phi$ play now the role of the dynamical fields. The background
equations should not be taken into account before variation of
$\lag^{dyn}_D$ with respect to $\Bar g^a$ and $\Bar \Phi$. Here, we
need in the gravitational equations related to the Lagrangian
(\ref{lag}) only on a vacuum background ($\Bar \lag_{m} = 0$):
  \be
\hat G^{L}_{\mu\nu}  = \k\hat t_{\mu\nu}\,.
 \m{PERT-munu}
 \ee
They are equivalent to the equations in the usual form related to
the Lagrangian (\ref{lag-g}). The linear in $h^a$ expression and the
generalized symmetric energy-momentum of the perturbations $h^a$ and
$\phi$ are defined as
 \be
 \hat G^{L}_{\mu\nu} \equiv
 \frac{\delta }{\delta \Bar g^{\mu\nu}}
 h^a
\frac{\delta \Bar \lag_{g}}{\delta \Bar { g}^{a}}\,,\qquad \hat
t_{\mu\nu} \equiv 2\frac{\delta\lag^{dyn}_D}{\delta \Bar
 g^{\mu\nu}}\,.
 \m{GL-q}
 \ee
With taking into account the gravitational background equations one
can rewrite the left hand side of (\ref{PERT-munu}) with another
{\em independent} gravitational variables, instead of $h^a$:
 \be
 \hat l^{\mu\nu}_{(a)} \equiv
 h^a {{ \di \Bar {\hat g}^{\mu\nu}} / {\di \Bar g^a}}\,.
 \m{B40}
 \ee

Usually to construct superpotentials one contracts the left hand
side of (\ref{PERT-munu}) with the background Killing vectors and
provides direct algebraic transformations, like in
\cite{AbbottDeser82,DT2}. We present a more universal way
\cite{Petrov2008,Petrov2005b} as follows. Conserved quantities can
be obtained and described analyzing {\it only} the scalar density
 \be \lag_1 \equiv -\frac{1}{2\k} h^a
\frac{\delta \Bar \lag_{g}}{\delta \Bar { g}^{a}} \equiv
 -\frac{1}{2\k}\hat l^{\alf\beta}_{(a)} \frac{\delta \Bar
\lag_{g}}{\delta \Bar {\hat g}^{\alf\beta}}\, ,
 \m{Lag-1}
 \ee
which is the linear pure metric term in the Lagrangian (\ref{lag}).
After applying the N{\oe}ther procedure directly and some algebraic
transformations one obtains the superpotential:
  \be
\hat {\cal I}_{S}^{\alf\beta}  \equiv  {\textstyle{4\over 3}}\l(
 2\xi^\sig \BD_\lam  \hat N_{\sig}{}^{[\alf|\beta]\lam}   -
\hat N_{\sig}{}^{[\alf|\beta]\lam}
 \BD_\lam  \xi^\sig\r) \equiv {\textstyle{8\over 3}}
  \BD_\lam \l( \hat N_{\sig}{}^{[\alf|\beta]\lam}\xi^\sig\r)   -
4\hat N_{\sig}{}^{[\alf|\beta]\lam}
 \BD_\lam  \xi^\sig\,
 \m{(+16+A)}
 \ee
where
 \be
 \hat N^{\rho\lam|\mu\nu}  =   {\di \lag_1}/{\di \Bar
g_{\rho\lam,\mu\nu}}\,.
 \m{NL1}
 \ee
Conservation laws in the field-theoretical approach are based on the
equations (\ref{PERT-munu}) with the {\em symmetrical}
energy-momentum in (\ref{GL-q}), therefore we use the subscript
``${}_S$''.

To compare the expressions of this subsection with the known ones in
4D GR we again could do simplifications, like in previous
subsections. Then, the equations (\ref{PERT-munu}) in varibles $\hat
l^{\mu\nu} = \hat g^{\mu\nu} - \Bar{\hat g^{\mu\nu}}$ repeat the
equations presented by Weinberg \cite{Weinberg-book}, the
superpotential (\ref{(+16+A)}) again transforms into the Papapetrou
superpotential \cite{Papapetrou48}, note that in the 4D GR the
Belinfante and the field-theoretical approaches give the same result
\cite{Petrov2008}. Under more weak restrictions, say, to AdS/dS
backgrounds in 4D GR the superpotential (\ref{(+16+A)}) goes to the
Abbott-Deser expression \cite{AbbottDeser82}. At last, the
expression (\ref{(+16+A)}) generalizes both our superpotential in 4D
GR \cite{PK} and the Deser-Tekin expressions \cite{DT2} in quadratic
theories, if there one considers equations not more than of the
second order.

\subsection{Instant remarks}
\m{R}

The first remark is of general type. Keeping in mind cosmological
and astrophysical applications, conservation laws of the type
(\ref{generalCLs}) and (\ref{charges}) could play an important role
to connect non-local conserved quantities (surface integrals) with
local ones (currents for perturbations).

The other remarks are as follow. 1) The canonical and Belinfante
corrected approaches are bimetric. After substituting the
decompositions (\ref{g-Dec}) into the superpotentials (\ref{IsupD})
and (\ref{(supIB+}) they are presented in the explicitly perturbed
form, like (\ref{(+16+A)}). 2) Here, we consider theories, equations
of which have derivatives not higher than of second order only, like
the Lovelock theories \cite{Lovelock}. However our results can be
easily generalized for theories with derivatives of higher orders
both in Lagrangians and in equations. Thus we agree with the
discussion in the paper \cite{Bouchareb} related to our results
\cite{Petrov2005b}. 3) Viewing all the three types of the
superpotentials (\ref{IsupD}), (\ref{(supIB+}) and (\ref{(+16+A)}),
it is important to note that their form is left the same both for
vacuum and for non-vacuum backgrounds, and both for Killing and for
arbitrary displacement vectors. 4) Unlike the canonical quantities,
the Belinfante corrected and the field-theoretical superpotentials
do not depend on divergences at all that could be necessary in
various situations. 5) All the presented here procedures give well
defined superpotentials in the sense that they are {\it uniquely}
defined by the Lagrangian.

\section{Superpotentials in the
EGB gravity}
 \m{EGBsuperpotentials}
\setcounter{equation}{0}

Here, applying the formulae of the above section we derive out
superpotentials in the EGB gravity presented by the action:
 \be
 S  = -\frac{1}{2\k} \int d^D x\lag_{EGB} = -\frac{1}{2\k}
 \int d^D x\sqrt{-g} \l[R - 2\Lambda_0 +
 \alpha\l(R^2_{\mu\nu\rho\sig} - 4 R^2_{\mu\nu} + R^2\r)\r]
 \,,
 \m{EGBaction}
 \ee
$\k = 2\Omega_{D-2}G_D> 0$  and $\alpha >0$; $G_D$ is the
$D$-dimensional Newton constant, and we restrict ourselves by
$\Lambda_0 \leq 0$. Below, the subscripts ``${}_{E}$'' is related to
the pure Einstein part in (\ref{EGBaction}), and the subscript
``${}_{GB}$'' is related to the Gauss-Bonnet part with the
connection constant $\alf$.

At first we calculate the canonical superpotential (\ref{IsupD}).
With the use of (\ref{DeltaDefD}) and (\ref{(17-DmD)}) we represent
$\lag_{EGB}$ into a covariantized form with the external background
metric. After that we calculate the coefficients (\ref{(+4+)}) and
(\ref{(+5+)}), substitute them into (\ref{+IsupD+}) and obtain
 \bea
\hat \imath^{\alf\beta}_C  &= & {}_{E}\hat \imath^{\alf\beta}_C
+{}_{GB}\hat \imath^{\alf\beta}_C
 \nonumber\\ &=&
{1\over \k}\big({\hat g^{\rho[\alf}\Bar D_{\rho} \xi^{\beta]}} +
\hat g^{\rho[\alf}\Delta^{\beta]}_{\rho\sig}\xi^\sig\big)
 \nonumber\\ &+ & \frac{\alf\sqrt{-g}}{\k}
 \l\{6 R_{\tau}{}^{[\alf\beta]\rho}\Delta^{\tau}_{\sig\rho}-
 2R_{\sig}{}^{\rho\tau[\alf}\Delta^{\beta]}_{\rho\tau}  -
 6 R^{\rho[\alf\beta]}{}_\sig\Delta^{\tau}_{\tau\rho} +
 4 R^{\rho[\alf}\Delta^{\beta]}_{\sig\rho} -
 2 R^{\rho}_{\sig}\Delta^{[\alf}_{\rho\tau}g^{\beta]\tau}\r.
 \nonumber\\&-& \l.
 4 R^{\rho\tau}\delta^{[\alf}_{\sig}\Delta^{\beta]}_{\rho\tau} +
 2 R^{\rho[\alf}g^{\beta]\tau}\Delta^{\pi}_{\rho\tau}g_{\sig\pi} +
 8 R^{[\alf}_{\rho}g^{\beta]\tau}\Delta^{\rho}_{\sig\tau} +
 8 R^{[\alf}_{\sig}g^{\beta]\rho}\Delta^{\tau}_{\tau\rho} +
 4 R^{[\alf}_{\sig}\Delta^{\beta]}_{\tau\rho}g^{\tau\rho} +
  4 R^{[\alf}_{\sig}g^{\beta]\rho}\Delta^{\tau}_{\tau\rho} \r.
 \nonumber\\&-& \l. 2 R \Delta^{[\alf}_{\sig\rho}g^{\beta]\rho}-
  4 R \Delta^{\tau}_{\tau\rho}\delta^{[\alf}_{\sig}g^{\beta]\rho} +
 \BD_\tau R_{\sig}{}^{\tau\alf\beta} + 6 g_{\sig\rho}g^{\tau[\alf}\BD_\tau R^{\beta]\rho} +
 2\delta^{[\alf}_\sig g^{\beta]\rho}\BD_\rho R\r\}\xi^\sig
 \nonumber\\&-&
\frac{2\alf\sqrt{-g}}{\k}\l\{R_\sig{}^{\lam\alf\beta} +
 4 g^{\lam[\alf}R^{\beta]}_\sig + \delta_\sig^{[\alf}g^{\beta]\lam}R
 \r\}\BD_\lam \xi^\sig\, .
 \m{CanSupEGB}
 \eea
The barred expression is derived from the above and is simpler
because $\Bar {\Delta^\gamma}_{\alf\beta}\equiv 0$:
 \bea
\Bar{\hat \imath^{\alf\beta}_C}  &= & {}_{E}\Bar{\hat
\imath^{\alf\beta}_C} +{}_{GB}\Bar{\hat \imath^{\alf\beta}_C}
 \nonumber\\ &=&
{1\over \k} \Bar {D^{[\alf} \hat\xi^{\beta]}}
 \nonumber\\ &+ & \frac{\alf}{\k}
 \l\{
 \Bar{D_\tau \hat R}_{\sig}{}^{\tau\alf\beta} +
 6 \Bar{D^{[\alf} \hat R^{\beta]}_\sig} +
 2\delta^{[\alf}_\sig \Bar{D^{\beta]}\hat R}\r\}\xi^\sig
 \nonumber\\&-&
\frac{2\alf}{\k}\l\{\Bar{\hat R}_\sig{}^{\lam\alf\beta} +
 4 \Bar{g^{\lam[\alf}\hat R^{\beta]}_\sig} + \delta_\sig^{[\alf}
 \Bar{g^{\beta]\lam}\hat R}
 \r\}\BD_\lam \xi^\sig\, .
 \m{BarCanSupEGB}
 \eea
To finalize constructing the superpotential (\ref{IsupD}) one needs
to fix a divergence in the Lagrangian (\ref{ArbitraryLagPert}).  As
we promised, we consider two possibilities. At first, we follow the
recommendation in \cite{DerKatzOgushi}:
 $\hat d^\lam_{DKO} =
 -2\k{\hat n}_{\sig}{}^{\lam\alf\beta}\Delta^\sig_{\alf\beta}$.
Thus,
 \bea
\hat d^\lam_{DKO} &=&  {}_{E}\hat d^\lam + {}_{GB}\hat d^\lam =
-2\k\l({}_{E}{\hat n}_{\sig}{}^{\lam\alf\beta}+ {}_{GB}{\hat
n}_{\sig}{}^{\lam\alf\beta}\r)\Delta^\sig_{\alf\beta} \nonumber\\
&=& {2\sqrt{-g}} \Delta^{[\alf}_{\alf\beta}\hat g^{\lam]\beta}  +
 4 \alf\l(
\hat R_\sig{}^{\alf\beta\lam} - 4\hat R^{[\alf}_\sig g^{\lam]\beta}+
\delta^{[\alf}_\sig g^{\lam]\beta} \hat
R\r)\Delta^\sig_{\alf\beta}\, .
 \m{divd}
 \eea
The Katz-Livshits \cite{KatzLivshits} approach leads to
 \bea
\hat d^\lam_{KL}  &=&  {}_{E}\hat d^\lam + {}_{GB}\hat d^\lam
\nonumber\\ &=& {2\sqrt{-g}} \Delta^{[\alf}_{\alf\beta}\hat
g^{\lam]\beta}  +
 4 \alf\l(
\hat R_\sig{}^{\alf\beta\lam} - 2\hat R^{[\alf}_\sig g^{\lam]\beta}
-2\delta^{[\alf}_\sig \hat R^{\lam]\beta}+ \delta^{[\alf}_\sig
g^{\lam]\beta} \hat R\r)\Delta^\sig_{\alf\beta}\, .
 \m{divdKL}
 \eea
As is seen, in GR case both of these lead to the choice of KBL.
Keeping in mind (\ref{CanSupEGB}) - (\ref{divdKL}), we derive the
superpotential in the canonical prescription (\ref{IsupD}) for the
EGB gravity:
 \bea
 \hat {\cal I}^{\alf\beta}_{C} &=&  {}_{E}\hat {\cal I}^{\alf\beta}_{C}
+{}_{GB}\hat {\cal I}^{\alf\beta}_{C}\nonumber\\ &=&
 {1\over \k}\l({\hat g^{\rho[\alf}\Bar D_{\rho} \xi^{\beta]}} +
\hat g^{\rho[\alf}\Delta^{\beta]}_{\rho\sig}\xi^\sig-  \Bar
{D^{[\alf} \hat\xi^{\beta]}} + \xi^{[\alf}{}_{E}\hat
d^{\beta]}\r)\nonumber\\
&+& {}_{GB}{\hat \imath^{\alf\beta}_C} - {}_{GB}\Bar{\hat
\imath^{\alf\beta}_C} + \k^{-1}\xi^{[\alf}{}_{GB}\hat d^{\beta]}\,
 \m{CanEGB}
 \eea
where ${}_{GB}\hat d^\alf$ is derived from (\ref{divd}) or
(\ref{divdKL}). The full Einstein part ${}_{E}\hat {\cal
I}^{\alf\beta}_{C}$ is exactly the KBL superpotential both in 4D GR
\cite{KBL} and in multi-D GR \cite{DerKatzOgushi,KatzLivshits}.

Now let us turn to the Belinfante corrected superpotential
(\ref{(supIB+}) in EGB gravity. Substituting the correspondent
coefficients (\ref{(+5+)}) into (\ref{+(supIB+}) one obtains
 \bea
\hat \imath^{\alf\beta}_{B}  &= & {}_{E}\hat \imath^{\alf\beta}_{B}
+{}_{GB}\hat \imath^{\alf\beta}_{B}
 \nonumber\\ &=&
{1\over \k}\l[\l( \delta^{[\alf}_\sig \BD_\lam \hat g^{\beta]\lam}
-\BD^{[\alf}\hat g^{\beta]\rho}\Bar g_{\rho\sig} \r)\xi^\sig +\hat
g^{\lam[\alf}\BD_\lam \xi^{\beta]}\r]\nonumber\\
&+&{\alf\over \k}\BD_\lam\l\{\hat R_\sig{}^{\lam\alf\beta} +
 4 g^{\lam[\alf}\hat R^{\beta]}_\sig
 +\l[2\hat R_\tau{}^{\rho\lam[\alf}
-2\hat R^{\rho\lam}{}_\tau{}^{[\alf} - 8\hat R^\lam_\tau
g^{\rho[\alf}\r.\r.
 \nonumber\\&+& \l.\l.4
\hat R^\rho_\tau g^{\lam[\alf} +4 g^{\rho\lam} \hat R^{[\alf}_\tau +
2\hat R\l( \delta^\lam_\tau g^{\rho[\alf}- \delta^\rho_\tau
g^{\lam[\alf}\r)\r]\Bar g^{\beta]\tau}\Bar g_{\rho\sig} \r\}\xi^\sig
 \nonumber\\&-&{2\alf\over \k}\l\{{\hat R}_\sig{}^{\lam\alf\beta} +
 4 {g^{\lam[\alf}\hat R^{\beta]}_\sig} + \delta_\sig^{[\alf}
 g^{\beta]\lam}\hat R
 \r\}
 \BD_\lam \xi^\sig\,.
 \m{BelSupEGB}
 \eea
The barred expression is significantly simpler
 \bea
&{}&\Bar{\hat \imath^{\alf\beta}_{B}}  =  {}_{E}\Bar{\hat
\imath^{\alf\beta}_{B}} +{}_{GB}\Bar{\hat \imath^{\alf\beta}_{B}}
 \nonumber\\ &=&
{1\over \k}\Bar{\hat g^{\lam[\alf}D_\lam \xi^{\beta]}}-{2\alf\over
\k}\l\{\Bar{\hat R}_\sig{}^{\lam\alf\beta} +
 4 \Bar{g^{\lam[\alf}\hat R^{\beta]}_\sig} + \delta_\sig^{[\alf}
 \Bar{g^{\beta]\lam}\hat R}
 \r\}
 \BD_\lam \xi^\sig\,.
 \m{BarBelSupEGB}
 \eea
 Thus, keeping in mind (\ref{BelSupEGB}) and  (\ref{BarBelSupEGB}), we derive the
superpotential in the Belinfante corrected prescription
(\ref{(supIB+}) for the EGB gravity:
 \bea
 \hat {\cal I}^{\alf\beta}_{B} &=&  {}_{E}\hat {\cal I}^{\alf\beta}_{B}
+{}_{GB}\hat {\cal I}^{\alf\beta}_{B}\nonumber\\ &=&
 {1\over \k}\l( \xi^{[\alf} \BD_\lam \hat l^{\beta]\lam}
-\BD^{[\alf}\hat l^{\beta]}_{\sig}\xi^\sig  +\hat
l^{\lam[\alf}\BD_\lam \xi^{\beta]}\r)+ {}_{GB}{\hat
\imath^{\alf\beta}_{B}} - {}_{GB}\Bar{\hat \imath^{\alf\beta}_{B}}
\,
 \m{BelEGB}
 \eea
where $\hat l^{\alf\beta}= \hat g^{\alf\beta}- \Bar{\hat
g^{\alf\beta}}$. The Einstein part ${}_{E}\hat {\cal
I}^{\alf\beta}_{B}$, being constructed in arbitrary $D$ dimensions,
has exactly the form of the Belinfante corrected superpotential in
the 4D GR \cite{PK}.

At last, turn to the symmetrical  superpotential (\ref{(+16+A)}) in
the framework of the EGB case.  To concretize a calculation and to
have a possibility to compare with \cite{DT2}, we define the
perturbations from the set (\ref{g-Dec}) as
 $
h^a = g_{\alf\beta} - \Bar g_{\alf\beta}=h_{\alf\beta}\, .
 $
Thus the Lagrangian in (\ref{Lag-1}) has to be calculated as
 $
\lag_1 = -({2\k})^{-1}h_{\alf\beta} {\delta \Bar \lag_{EGB}}/{\delta
\Bar g_{\alf\beta}}
 $.
Calculating on this basis the coefficients (\ref{NL1}) one has
 \bea
&{}& N^{\rho\lam|\mu\nu}= N^{\rho\lam|\mu\nu}_{E} +
N^{\rho\lam|\mu\nu}_{GB} = \nonumber\\& -&\frac{1}{4\k}\l[\Bar
g^{\rho\lam}h^{\mu\nu}+ \Bar g^{\mu\nu}h^{\rho\lam} -\Bar
g^{\rho(\mu}h^{\nu)\lam}-\Bar g^{\lam(\mu}h^{\nu)\rho}+
h^\sig_\sig\l(\Bar g^{\rho(\mu}\Bar g^{\nu)\lam}-  \Bar
g^{\rho\lam}\Bar g^{\mu\nu}\r)\r]\,
 \nonumber
\\& - & \frac{\alpha}{2\k}\l\{h^\sig_\sig\l[\l(\Bar
g^{\rho(\mu}\Bar g^{\nu)\lam} - \Bar g^{\rho\lam}\Bar
g^{\mu\nu}\r)\Bar R - 2\l(\Bar g{}^{\rho(\mu}\Bar R{}^{\nu)\lam}+
\Bar R{}^{\rho(\mu}\Bar g{}^{\nu)\lam} \r) + 2\l(\Bar
g{}^{\rho\lam}\Bar R{}^{\mu\nu}+ \Bar
R{}^{\rho\lam}\Bar g{}^{\mu\nu}\r)\r.\r.\nonumber\\
& + &\l. 2 \Bar R^{\rho(\mu\nu)\lam}\r]+\l(\Bar
g^{\rho\lam}h^{\mu\nu}+ \Bar g^{\mu\nu}h^{\rho\lam} -\Bar
g^{\rho(\mu}h^{\nu)\lam}-\Bar g^{\lam(\mu}h^{\nu)\rho} \r)\Bar R +
2\l(h{}^{\rho(\mu}\Bar R{}^{\nu)\lam}+ \Bar R{}^{\rho(\mu}
h{}^{\nu)\lam}\r)\nonumber\\& -& 2\l(h{}^{\rho\lam}\Bar
R{}^{\mu\nu}+\Bar R{}^{\rho\lam}h{}^{\mu\nu} \r) -4\l(\Bar
g^{\rho\lam}\Bar R{}_\sig^{(\mu}h{}^{\nu)\sig}+ \Bar g^{\mu\nu}\Bar
R{}_\sig^{(\rho}h{}^{\lam)\sig}\r) + 4\l(\Bar R{}_\sig^{(\rho}\Bar
g^{\lam)(\mu}h^{\nu)\sig} +
\Bar R{}_\sig^{(\mu}\Bar g^{\nu)(\rho}h^{\lam)\sig}\r) \nonumber\\
&-& 2\l(\Bar g^{\rho(\mu}\Bar g^{\nu)\lam} - \Bar g^{\rho\lam}\Bar
g^{\mu\nu}\r)\Bar R{}_\sig^\tau h{}^\sig_\tau - 4\l(\Bar
R{}_\sig{}^{(\rho\lam)(\mu}h{}^{\nu)\sig}+ \Bar
R{}_\sig{}^{(\mu\nu)(\rho}h{}^{\lam)\sig} \r)\nonumber\\
&+& \l.2h_{\sig\tau}\l(\Bar R{}^{\sig\mu\tau(\rho}\Bar g^{\lam)\nu}
+ \Bar R{}^{\sig\nu\tau(\rho}\Bar g^{\lam)\mu}\r) + 2h_{\sig\tau}\l(
\Bar g^{\rho\lam}\Bar R^{\sig\mu\nu\tau} + \Bar g^{\mu\nu}\Bar
R^{\sig\rho\lam\tau}\r)\r\}\,.
  \m{NNN}
 \eea
Substituting them into (\ref{(+16+A)}) one obtains
  \bea
\hat {\cal I}_{S}^{\alf\beta} & =& {}_{E}\hat {\cal
I}_{S}^{\alf\beta} + {}_{GB}\hat {\cal I}_{S}^{\alf\beta}
\nonumber\\& =& \frac{\sqrt{-\Bar g}}{\k}
 \l( \xi_\nu
\Bar D^{[\alf}h^{\beta]\nu}- \xi^{[\alf} \Bar D_\nu h^{\beta]\nu} +
\xi^{[\alf} \Bar D^{\beta]}h - h^{\nu[\alf}\Bar D_\nu\xi^{\beta]}
+\half h \Bar D^{[\alf} \xi^{\beta]}\r)
 \nonumber\\& +& {{4\over 3}}\l(
 2\xi_\sig \BD_\lam  \hat N_{{GB}}^{\sig[\alf|\beta]\lam}   -
\hat N_{{GB}}^{\sig[\alf|\beta]\lam}
 \BD_\lam  \xi_\sig\r)\,.
 \m{SymEGB-h}
 \eea
This expression coincides with the Deser-Tekin superpotential
\cite{DT2} if one chooses the EGB gravity on the AdS background. The
expression (\ref{SymEGB-h}) calculated for $h^a=h_{\alf\beta}$ is a
particular case of the more general case with the decomposition
(\ref{g-Dec}) and the redefinition (\ref{B40}):
 \bea
\hat {\cal I}_{S}^{\alf\beta} & =& {}_{E}\hat {\cal
I}_{S}^{\alf\beta} + {}_{GB}\hat {\cal I}_{S}^{\alf\beta} =
 \frac{1}{\k}
 \l( \hat
l^{\sig[\mu}_{(a)}\Bar D_\sig\xi^{\rho]}+ \xi^{[\mu}\Bar D_\sig \hat
l^{\rho]\sig}_{(a)}-\bar D^{[\mu}\hat l^{\rho ]}_{(a)\sig}
\xi^\sig\r)
 \nonumber\\& +& {{4\over 3}}\l(
 2\xi_\sig \BD_\lam  \hat N_{{GB}}^{\sig[\alf|\beta]\lam}(\hat l_{(a)})   -
\hat N_{{GB}}^{\sig[\alf|\beta]\lam}
 \BD_\lam (\hat l_{(a)})  \xi_\sig\r)\,.
 \m{SymEGB-la}
 \eea

Of course, the above defined superpotentials differs one from
another. However, returning to 4D GR, as we remarked in
Introduction, all the three approaches satisfy the main tests.
Indeed, calculating at infinity the surface integrals
(\ref{charges}) for isolated systems in 4D GR \cite{Petrov2008}, one
obtains the same accepted results with using each of the Einstein
parts in the superpotentials (\ref{CanEGB}), (\ref{BelEGB}) and
(\ref{SymEGB-la}). In the case of asymptotically flat spacetimes at
{\em spatial infinity} the situation is more simple: all the
differences between these kinds of superpotentials do not contribute
into (\ref{charges}) \cite{PK,Petrov2008}. Analogously, in the next
sections, we show  that all the three approaches give the standard
mass for the S-AdS BH in EGB gravity.

A more complicated case is an isolated system at {\em null
infinity}. In 4D GR both the canonical and Belinfante corrected
approaches give the same result \cite{PK} coinciding with the
standard Bondi-Sachs energy-momentum flux \cite{BMS}. Another
situation is in the symmetrical approach. A different choice of
variables from (\ref{(1)}) and respectively a different
decompositions (\ref{g-Dec}) lead to different $h^a$. Then variables
$\hat l^{\mu\nu}_{(a)}$ in (\ref{B40}) differ one from other in the
second order in perturbations. This difference is explicitly
incorporated into the left hand side of (\ref{PERT-munu}), and
respectively into the superpotentials (\ref{(+16+A)}), or
(\ref{SymEGB-la}) in the EGB case. Already in \cite{BMS} it has been
remarked that the difference in the second order is important in a
calculation for a radiating isolated system in 4D GR. It turns out
\cite{Petrov2008} that only the choice
 $h^a = \hat g^{\mu\nu} - \Bar {\hat
g}^{\mu\nu}\equiv \hat l^{\mu\nu}\,
 $
gives the standard result \cite{BMS} (not another, including
$h^a=h_{\alf\beta}$). Concerning EGB gravity, a) the different $\hat
l^{\mu\nu}_{(a)}$ in (\ref{B40}) give the same standard mass for the
S-AdS BH calculated below, and b) we only plan to consider radiating
systems in future.

Considering different variables $\hat l^{\mu\nu}_{(a)}$ in
(\ref{B40}), we note also that only for $h^a=\hat l^{\alf\beta}$ the
Einstein parts in (\ref{BelEGB}) and (\ref{SymEGB-la}) coincide:
${}_{E}\hat {\cal I}_{B}^{\mu\nu}={}_{E}\hat {\cal I}_{S}^{\mu\nu}$
that could be interpreted as an advantage of the choice $h^a=\hat
l^{\alf\beta}$. However, in an arbitrary gravitation theory $\hat
{\cal I}_{B}^{\mu\nu}\neq \hat {\cal I}_{S}^{\mu\nu}$, compare the
Gauss-Bonnet parts in (\ref{BelEGB}) and (\ref{SymEGB-la}).
Nevertheless, because many multi-D gravities, as a rule, are the
Einstein theories with corrections, one again could prefer $\hat
l^{\mu\nu}$ from the set $\hat l^{\mu\nu}_{(a)}$ comparing the
Einstein parts {\em only}.

\section{Static spherically symmetric solutions}
 \m{Ssss}
 \setcounter{equation}{0}

Many interesting solutions in the vacuum EGB gravity have the
Schwarzschild-like form:
  \be
 d s^2 = -fdt^2 +  f^{-1}dr^2 +
 r^2\sum_{a,b}^{D-2}q_{ab}dx^adx^b\,
 \m{S-AdS}
 \ee
where $f= f(r)$, the last term describes $(D-2)$-dimensional sphere
of the radius $r$, and $q_{ab}$ depends on coordinates on the sphere
only. As a background we choose again the solution of the {\em same}
form:
 \be
 d\Bar s^2 = -\Bar fdt^2 + (\Bar f)^{-1}dr^2 +
 r^2\sum_{a,b}^{D-2}q_{ab}dx^adx^b\, .
 \m{AdS}
 \ee
In this section, we present the general formulae, which can be used
for calculating the mass of the perturbed system (\ref{S-AdS}) with
respect to the background (\ref{AdS}). Due to a spherical symmetry,
for calculating conserved quantities one needs only in the
$01$-component of superpotentials in (\ref{charges}). We also note
that a background of the type (\ref{AdS}) has the Killing vector
 $
 \lam^\alf =\{-1,\,\bf 0\},
 $
which is just necessary for calculating a mass of the system.

At first, we calculate the  component $\hat {\cal I}^{01}_{C}$ of
the canonical superpotential (\ref{CanEGB}). We use the metrics
(\ref{S-AdS}) and (\ref{AdS}) directly in (\ref{CanSupEGB}) and
(\ref{divd}) with $\xi^\alf = \lam^\alf$. Thus,
 \bea
 {\hat\imath}^{01}_C &+& \k^{-1}\lam^{[0}\hat d^{1]}_{DKO}
 = \frac{\sqrt{-\Bar g}}{2\k r}\l[\frac{r\Bar{f}'}{2}\l(\frac{f}{\Bar f}+\frac{\Bar
f}{f} \r) - (f-\Bar f)(D-2)\r]\nonumber \\&-&
 \frac{\alf\sqrt{-\Bar g}}{\k r^2}(D-2)\l[f\l(f'- rf''\r)-\Bar f f'(D-3) \r] \nonumber \\
& -&\frac{\alf\sqrt{-\Bar g}}{\k r^3}(D-2)(D-3)(f-1)
\l[\frac{r\Bar{f}'}{2}\l(\frac{f}{\Bar f}+\frac{\Bar f}{f} \r)
-(f-\Bar f)(D-2)+2\Bar f  \r].
  \m{Can-i+d}
 \eea
We have used $\sqrt{- g}=\sqrt{-\Bar g}$. The 01-component
calculated with the divergence (\ref{divdKL}) is
 \bea
 {\hat\imath}^{01}_C &+& \k^{-1}\lam^{[0}\hat d^{1]}_{KL}
 = \frac{\sqrt{-\Bar g}}{2\k r}\l[\frac{r\Bar{f}'}{2}\l(\frac{f}{\Bar f}+\frac{\Bar
f}{f} \r) - (f-\Bar f)(D-2)\r]\nonumber \\&-&
 \frac{\alf\sqrt{-\Bar g}}{\k r^2}(D-2)\l[\Bar f\l(f'- rf''\r)+(f-2\Bar f) f'(D-3) \r] \nonumber \\
& -&\frac{\alf\sqrt{-\Bar g}}{\k r^3}(D-2)(D-3)(f-1)
\l[\frac{r\Bar{f}'}{2}\l(\frac{f}{\Bar f}+\frac{\Bar f}{f} \r)
-(f-\Bar f)D+2\Bar f  \r].
 \m{Can-i+dKL}
 \eea
The background expression for both the cases (\ref{Can-i+d}) and
(\ref{Can-i+dKL}) is the unique one
 \bea
\Bar{\hat\imath^{01}_C} & = &\frac{\sqrt{-\Bar g}}{2\k }\Bar{f}' +
 \frac{\alf\sqrt{-\Bar g}}{\k r^2}(D-2)\Bar f\l[r\Bar f''+ \Bar
 f'\l(D-4\r)\r]
 \nonumber \\
& -&\frac{\alf\sqrt{-\Bar g}}{\k r^3}(D-2)(D-3)(\Bar f-1)
\l(r\Bar{f}'+ 2\Bar f\r)\,.
  \m{BarCan-i+d}
 \eea
Of course, it is also calculated directly from (\ref{BarCanSupEGB}).
The component $\hat {\cal I}^{01}_{C}$ in (\ref{CanEGB}) is obtained
after substraction of (\ref{BarCan-i+d}) from (\ref{Can-i+d}) or
(\ref{Can-i+dKL}).

Now, calculate the component $\hat {\cal I}^{01}_{B}$  of the
Belinfante corrected superpotential (\ref{BelEGB}) for the relative
systems (\ref{S-AdS}) and (\ref{AdS}) and the displacement vector
$\xi^\alf = \lambda^\alf$. For (\ref{BelSupEGB}) one has
 \bea
 \hat\imath^{01}_{B}
& =& \frac{\sqrt{-\Bar g}}{2\k
r}\l[\frac{r\Bar{f}'}{2}\l(3\frac{f}{\Bar f}-\frac{\Bar f}{f}
\r)-rf' \l(1-\frac{\Bar f^2}{f^2} \r)- (f-\Bar f)(D-2)\r]\nonumber
\\&-&
 \frac{\alf\sqrt{-\Bar g}}{\k r^2}(D-2)\l[\Bar f\l(f'- rf''\r)
 \l(1-\frac{\Bar f}{f} \r)- f f'\l(1-\frac{\Bar f^2}{f^2} \r)(D-3) \r] \nonumber \\
& +&\frac{\alf\sqrt{-\Bar g}}{\k r^3}(f-1)(D-2)(D-3)
\l[\frac{r\Bar{f}'}{2}\l(\frac{\Bar f}{f}-3\frac{f}{\Bar f} \r)+
(rf'- 2f)
\l(1-\frac{\Bar f^2}{f^2} \r)\r. \nonumber \\
 &+&\l.(f-\Bar f)(D-2)\r].
  \m{Bel-i}
 \eea
The barred expression (\ref{Bel-i}) is
 \be
\Bar{\hat\imath^{01}_{B}} =\frac{\sqrt{-\Bar g}}{2\k }\Bar{f}'-
\frac{\alf\sqrt{-\Bar g}}{\k r^2}(D-2)(D-3) \Bar{f}'(\Bar f-1)\,.
  \m{BarBel-i}
 \ee
Of course, it is also calculated directly from (\ref{BarBelSupEGB}).
The component $\hat {\cal I}^{01}_{B}$ in (\ref{BelEGB}) is obtained
after subtraction of (\ref{BarBel-i}) from (\ref{Bel-i}).

At last, we calculate the $01$-component of the symmetrical
superpotential. For the metric perturbations of (\ref{S-AdS}) with
respect to (\ref{AdS}) one has only the non-zero components
 $
 h_{00}= -(f-\Bar f),~~h_{11}= -{(f-\Bar f)}/{f\Bar f}\,.
 $
Using them, $\xi^\alf = \lambda^\alf$ and (\ref{AdS}) in
(\ref{SymEGB-h}) directly one obtains
  \be
 \hat{\cal I}^{01}_{S}
 = \frac{\sqrt{-\Bar g}}{2\k r}(D-2)(f-\Bar f)\frac{\Bar f}{f}
 \l[-1 + 2\alf(\Bar f - 1)\frac{(D-3)(D-4)}{r^2}\r].
  \m{Sym-i}
 \ee
This expression is significantly simpler than corresponding
expressions in the other approaches. Besides,  being proportional to
$f-\Bar f$, it directly describes the perturbed system. At last, we
note that all the final expressions of this section are exact (not
approximate).

\section{The mass of the Schwarzschild-AdS black hole}
 \m{BHinEGB}
\setcounter{equation}{0}

As an example of (\ref{S-AdS}), we consider the S-AdS solution
\cite{BD+}:
 \be
 f(r) = 1 -\frac{r^2\Lambda'}{(D-2)(D-1)}
\l\{1 \pm \sqrt{1 - \frac{4\Lambda_{0}}{\Lambda'} - \frac{2
(D-2)(D-1)}{\Lambda'}\frac{\mu}{r^{D-1}}} \r\}\,
 \label{39bis}
 \ee
where $\mu$ is a constant of integration, and
 $
 \Lambda' = -
{(D-2)(D-1)}/{{2\alpha}(D-4)(D-3)}\,
 $
is defined {\em only} by the Gauss-Bonnet term. For the sake of
simplicity (to exclude numerous nuances) we restrict ourselves to $D
\geq 5$. As the important backgrounds we consider the AdS spacetime.
It is defined by (\ref{39bis}) at $\mu =0$ with the  effective
cosmological constant $ \Lambda_{eff}$:
 \be
\Bar f(r) = 1- r^2\frac{2\Lambda_{eff}}{(D-1)(D-2)}\,;\qquad
 \Lambda_{eff} = \frac{\Lambda'}{2} \l(1\pm \sqrt{1 -
 \frac{4\Lambda_{0}}{\Lambda'}}\r)\,.
 \m{fBar}
 \ee
In the linear approximation the perturbation (\ref{39bis}) with
respect to (\ref{fBar}) is
 \be
 \Delta f =f(r) - \Bar f(r)=  \pm \l(\sqrt{1 - \frac{4\Lambda_{0}}{\Lambda'}}\r)^{-1}
\frac{\mu}{r^{D-3}}\, .
 \m{Deltaf}
 \ee
We assume $1-{4\Lambda_{0}}/{\Lambda'} \neq 0$, the situation
$1-{4\Lambda_{0}}/{\Lambda'} = 0$ is discussed in the last section.

With a desire to consider the solution (\ref{39bis}) as a black hole
solution one has to choose the ``$-$'' sign (lower sign) because
only then one can define a horizon radius $r_+$ of the black hole
setting $f =0$. Besides, there is a qualitative difference between
the cases $D\geq 6$ and $D=5$.\m{gap} In the first case, vanishing
the constant of integration $\mu \goto 0$ corresponds vanishing the
horizon $r_+ \goto 0$. Thus, $\mu$ readily can be interpreted as a
mass parameter $M=\mu$, and the AdS solution (\ref{fBar}) can be
interpreted as a natural background for such black holes. Then the
asymptotic perturbations (\ref{Deltaf}) look quite natural. In the
case $D=5$, the situation is different: the horizon $r_+ \goto 0$
vanishes if $\mu \goto \mu_0$ where $\mu_0 = \alpha (D-3)(D-4) = 2
\alpha = -6/\Lambda'$. Then one can define a mass parameter as
$M=\mu-\mu_0$, for which again $r_+ \goto 0$ at $M \goto 0$. Thus
for such a black hole in 5 dimensions it is natural to choose a
vacuum background at $M = 0$ in (\ref{39bis}):
 \be
\Bar f(r) = 1 -\frac{r^2\Lambda'}{12} \l\{1 - \sqrt{1 -
\frac{4\Lambda_{0}}{\Lambda'} + \l(\frac{12}{r^2\Lambda'}\r)^2}
\r\}\,.
 \label{5DBHback}
 \ee
Only for negative $M=-\mu_0$ one approaches the AdS background with
$\Bar f$ in (\ref{fBar}), $\mu_0$ is called as a gap between the AdS
spacetime and a real black hole vacuum \cite{Cai}. Then, for $D=5$,
the perturbation with respect to (\ref{5DBHback}) asymptotically is
   \be
\Delta  f(r) = f(r) - \Bar f(r)= -\frac{M}{r^2} \l( \sqrt{1 -
\frac{4\Lambda_{0}}{\Lambda'}}\r)^{-1} \l[1 +
\frac{6}{r^4\Lambda'}\l(M- \frac{12}{\Lambda'}\r) \r] \,.
 \label{Deltaf5BH-lin}
 \ee
Here, the main order coincides with the order of a difference
between (\ref{5DBHback}) and (\ref{fBar}), therefore we conserve the
next order.

To calculate the mass of the S-AdS BH (\ref{39bis}) we use the
integral (\ref{charges}) under the requirement $r\goto \infty$. At
first we turn to the canonical prescription for both the cases
(\ref{Can-i+d}) and (\ref{Can-i+dKL}). We consider the linear
approximation of $\hat {\cal I}^{01}_{C}$ both in the perturbation
$\Delta f$ in (\ref{Deltaf}) with respect to (\ref{fBar}), $D\ge 6$,
and in the perturbation $\Delta f$ in (\ref{Deltaf5BH-lin}) with
respect to (\ref{5DBHback}), $D=5$. The {\em unique} expression can
be derived in all the cases because {\em only} the main orders in
(\ref{fBar}) and (\ref{Deltaf}),  or in (\ref{5DBHback}) and
(\ref{Deltaf5BH-lin}), contribute into the surface integral
(\ref{charges}). The asymptotic relations $(\Delta f)' =
-(D-3)\Delta f/r$, $~~(\Bar f)' = 2\Bar f/r$ are used. Then the
linear expression is
 \be
 \hat {\cal I}^{01}_{C} = -\frac{\sqrt{-\Bar g}}{2\k r}\Delta f (D-2)
 + \frac{\alf \sqrt{-\Bar
g}}{\k r^3}\Delta f \Bar f(D-2)(D-3)(D-4)\,.
 \m{CanEGB-lin}
 \ee
Substituting the main order from (\ref{fBar}) or (\ref{5DBHback}) we
obtain
 \be
 \hat {\cal I}^{01}_{C} = -\frac{\sqrt{-\Bar g}}{2\k r}\Delta f
 (D-2)\sqrt{1 -
 \frac{4\Lambda_{0}}{\Lambda_{EGB}}}\,\, \, .
 \m{CanEGB-lin+}
 \ee
Thus, for both the cases (\ref{Deltaf}) and (\ref{Deltaf5BH-lin}) we
have finally in the canonical approach:
 \be
 \hat {\cal I}^{01}_{C} = \frac{\sqrt{-\Bar g}}{2\k r} \frac{M}{r^{D-3}}(D-2).
 \m{CanEGBsup01}
 \ee
Substituting it into (\ref{charges}) together with
 $\sqrt{-\Bar g} = r^{D-2}\sqrt{\det q_{ij}}=r^{D-2}
\Omega_{D-2}$ one gets:
 \be
 E = (D-2)\frac{M}{4G_D}\,
 \label{E}
 \ee
that is the accepted result obtained with using the various
approaches (see \cite{BHS} - \cite{Okuyama,DerKatzOgushi,DT2} and
references therein). Recall also that both the cases (\ref{Can-i+d})
and (\ref{Can-i+dKL}) lead to (\ref{E}). Formally the expression
(\ref{CanEGB-lin+}) holds for $D \ge 5$. Thus, for $D = 5$ on AdS
background one can change $M \goto M+\mu_0$ in (\ref{CanEGBsup01})
and, respectively in (\ref{E}). This result could be interpreted as
an energy of the system presented by the BH together with the ``mass
gap'' $\mu_0$ on the AdS background. Then, indeed, one has to
calculate the energy of the 5D BH in the vacuum ``mass gap''.

To calculate the mass of the S-AdS BH (\ref{39bis}) with the use of
the Belinfante corrected expressions (\ref{Bel-i}) and
(\ref{BarBel-i}) we again consider the linear approximation and
follow all the same steps of the previous derivation. It turns out
that all the formulae and conclusions of the canonical prescription
from (\ref{CanEGB-lin}) are repeated exactly.

In the framework of the field-theoretical approach it is enough the
linear approximation:
 $
h_{00} = h^{11} = - \Delta f \,
 $
where $\Delta f$ can be defined both in (\ref{Deltaf}) and in
(\ref{Deltaf5BH-lin}). To calculate the mass of the S-AdS BH
(\ref{39bis}) we consider linear approximation of (\ref{Sym-i}) and
follow all the same steps of the previous approaches. Again, it
turns out that all the formulae and comments from (\ref{CanEGB-lin})
are repeated exactly.

At the above consideration, to stress an importance of the notion of
the ``mass gap'' in 5 dimensions and to show clearly a calculation
on the background of the ``mass gap'' vacuum, we are restricting to
the ``minus'' branch in (\ref{39bis}). However, the naive
interpretation of the mass permits to consider both the branches in
(\ref{39bis}) with the next conserving the two signs from
(\ref{fBar}) to (\ref{Deltaf5BH-lin}), like in
\cite{Paddila,Desercom,Petrov2005b}. One obtains again the accepted
quantity (\ref{E}).

\section{Discussion and concluding remarks} \m{Discussion}
 \setcounter{equation}{0}

There are infinitely many possibilities to construct conserved
quantities in metric theories. In the recent work \cite{Pitts} by
Pitts the very interesting idea is suggested, where  instead of many
complexes he considers any infinite-component object that is
conserved and that makes sense in every coordinate system/gauge.
This idea is in a convenience with our principal position, when all
the known possibilities to define conserved quantities are
important, if they are evidently non-contradictive and satisfy all
the acceptable tests.  Therefore we consider three approaches
simultaneously.

Here, we propagandize the derivation of perturbations in the
prescription of a field theory in the classical form. Considering a
background as an arena for perturbations, we connect conserved
charges with its symmetries expressed by Killing vectors. The
symmetrical approach is more illustrative because it derives
perturbations on backgrounds explicitly. The AdS background is more
popular therefore it is useful to consider just this case. We
substitute the metric (\ref{fBar}) into $N^{\rho\lam|\mu\nu}_{GB}$
in (\ref{NNN}) and obtain the relation:
 $N^{\rho\lam|\mu\nu}_{GB} = -{\l(1 \pm \sqrt{1 -
{4\Lambda_{0}}/{\Lambda'}}\r)} N^{\rho\lam|\mu\nu}_{E}
 $. After that
the symmetrical superpotential (\ref{(+16+A)}) in the EGB gravity
(\ref{SymEGB-h}) for of {\em arbitrary} type perturbations
$h_{\mu\nu}$ on the AdS background and with arbitrary $\xi^\alf$
acquires the form \cite{Petrov2005b}:
 \be
\hat {\cal I}_{S}^{\mu\rho} \equiv \pm \frac{\sqrt{-\Bar g}}{\k}
\sqrt{1 -
 \frac{4\Lambda_{0}}{\Lambda'}}
 \l(\xi^{[\mu} \Bar D_{\nu}h^{\rho]\nu}  - \xi_\nu
\Bar D^{[\mu}h^{\rho]\nu} - \xi^{[\mu} \Bar D^{\rho]}h -
h^{\nu[\mu}\Bar D^{\rho]}\xi_\nu - \half h \Bar D^{[\mu}
\xi^{\rho]}\r)\, .
 \m{DTsuperpotential}
 \ee
If the Killing vectors are used, then a) it is expressed through the
Abbott-Deser superpotential in the Einstein theory
\cite{AbbottDeser82,DT2}, $\hat I_{AD}^{\mu\rho}$, as $\hat {\cal
I}_{S}^{\mu\rho} = \mp \sqrt{1 - {4\Lambda_{0}}/{\Lambda'}}~ \hat
I_{AD}^{\mu\rho}$; b) the same expression (\ref{DTsuperpotential}),
note the signs, has been approached in \cite{Paddila,Desercom},
where the Deser-Tekin results \cite{DT2} were developed. Keeping in
mind the generalized metric perturbations (\ref{B40}), the
simplification of (\ref{SymEGB-la}) to the AdS background gives
 \be
 \hat {\cal I}_{S}^{\mu\rho} \equiv {1
\over \k}
 \l[{1}_{E} -{\l(1
\pm \sqrt{1 - \frac{4\Lambda_{0}}{\Lambda'}}\r)}_{GB}\r] \l(\hat
l^{\sig[\mu}_{(a)}\Bar D_\sig\xi^{\rho]}+ \xi^{[\mu}\Bar D_\sig \hat
l^{\rho]\sig}_{(a)}-\bar D^{[\mu}\hat l^{\rho ]}_{(a)\sig} \xi^\sig
\r)\,.
 \m{DTsuperpotential+}
 \ee
Thus, the superpotential (\ref{DTsuperpotential}) is the one of the
set (\ref{DTsuperpotential+}). The Einstein part in
(\ref{DTsuperpotential+}) formally coincides with the one in 4D GR
\cite{PK,Petrov2008}.

The superpotential (\ref{SymEGB-h}) can be thought as a
generalization of (\ref{DTsuperpotential}). Padilla, restricting his
own general results \cite{Paddila} to the AdS asymptotic, has found
out a consistence with the Deser-Tekin result
(\ref{DTsuperpotential}) also. We see that on the level of the AdS
background there is the consistence between the now discussed
results: of the present paper, Deser-Tekin's and Padilla's.
Returning to the three approaches presented here,  we have
demonstrated that they are powerful for calculations not only on AdS
backgrounds, but also on not maximally symmetrical backgrounds, like
(\ref{5DBHback}). Thus, concerning the criticism of the Deser and
Tekin approach in \cite{Paddila}, we show that developing their
method we are not restricted {\em only} by the AdS background.  But,
then it is interesting and important to compare the presented here
general approaches with the general Padilla approach \cite{Paddila}.
Of course, they are different: at least, our methods are Lagrangian
and are based on the conservation laws of the type
(\ref{generalCLs}), whereas in \cite{Paddila} surface integrals in
the Hamiltonian derivation play a main role. However, Padilla
applying his method, give the formula (129) in \cite{Paddila} for
calculating the mass of the model that we present in section 4: the
perturbed system of the type (\ref{S-AdS}) is considered with
respect to the background with the metric (\ref{AdS}). His integrand
in our notations could be rewritten as
 \be
 \hat{\cal I}^{01} = - \frac{\sqrt{-\Bar g}}{2\k r}\sqrt{f}
 \l\{6\l(\sqrt{f} - {\sqrt{\Bar f}}\r)+\frac{24\alf}{r^2}
 \l[\l(\sqrt{f} - {\sqrt{\Bar f}}\r)-
 {\textstyle \frac{1}{3}}\l(f\sqrt{f} - {\Bar f\sqrt{\Bar f}}\r) \r] \r\},
 \m{Padilla}
 \ee
which is quite different from each of the correspondent expressions
in our approaches (\ref{Can-i+d}) - (\ref{Sym-i}). Nevertheless,
substituting the perturbations (\ref{Deltaf5BH-lin}) on the
background (\ref{5DBHback}) into (\ref{Padilla}) one obtains again
the accepted result (\ref{E}) with $D=5$. Thus one can conclude that
the presented here methods are consistent with the Padilla approach
on the more complicated backgrounds than the AdS one.

The results of section \ref{BHinEGB} have been obtained under the
assumption that the factor \\$\sqrt{1 - 4\Lambda_0/\Lambda'}$ is not
equal to zero. Indeed, at the linear approximations the 01-component
of the superpotentials for calculating the mass in all the three
approaches is proportional to $\sqrt{1 - 4\Lambda_0/\Lambda'}$ (see
(\ref{CanEGB-lin+})). Thus, for
 $
 \Lambda'= 4\Lambda_{0}\,
 $
the global mass, probably, has to be treated as vanishing!? This
fact is remarked in the works \cite{Cai,Paddila,DerMor05,Desercom},
however without detailed discussion. Deruelle and Morisava
\cite{DerMor05} (canonical derivation), and Deser, Kanik and Tekin
\cite{Desercom} (field-theoretical derivation) have found out that
not only mass, but also angular momentum expressions for the
Kerr-AdS solution in EGB gravity have the same coefficient.

Considering the condition $\Lambda'= 4\Lambda_{0}$ authors, as a
rule, send readers to the paper \cite{Chamseddine2}, where the
situation is explained that gravitons do not propagate on AdS
backgrounds. Indeed, the linearized EGB vacuum equations calculated
as the left hand side in the equations (\ref{PERT-munu}) on the AdS
background has the same factor and disappear at the condition
$\Lambda'= 4\Lambda_{0}$. However, only in the framework of the
field-theoretical approach there is a direct connection between
vanishing the linear equations and the superpotentials (see
(\ref{DTsuperpotential+})). Inversely, in the canonical and the
Belinfante corrected derivations such a fact does not has a place:
the expressions (\ref{CanEGB}) and (\ref{BelEGB}) linearized around
the AdS background in {\em arbitrary} perturbations do not
proportional to $\sqrt{1 - 4\Lambda_0/\Lambda'}$. The
field-theoretical interpretation of the condition $\Lambda'=
4\Lambda_{0}$ looks also as a more suitable. Indeed, vanishing the
linear left hand side in (\ref{PERT-munu}) leads to vanishing the
total energy-momentum at the right hand side. For the case, when the
matter is absent, this means that the whole energy-momentum of the
{\em pure} metric perturbations is to be equal to zero. It is
absence of gravitons!

Difficulties connected with the condition $\Lambda'= 4\Lambda_{0}$
appear even if the linear approximation does not disappear. Thus, if
$\Lambda'= 4\Lambda_{0}$ our approaches do not give acceptable
results on the background of the ``mass gap'' in 5 dimensions.
Therefore, in future it is important to connect presented here
approaches with methods where the mass in this degenerated case is
defined acceptably; they are, e.g., the Regge and Teitelboim method
\cite{ReggeTeitelboim} in multi-$D$ application (see, e.g.,
\cite{BHS}) and the Paddila \cite{Paddila} Hamiltonian
constructions.

Keeping in mind a future development we note that the static
spherically symmetric Schwarzschild-like solutions could be examined
very effectively. However, the presented formalism is more
universal. Its methods can be especially transparent also for the
cases both of rotating black hole solutions (see, e.g., \cite{Aliev}
- \cite{Aliev2}) and of radiating solutions (see, e.g.,
\cite{GhoshDawood} - \cite{MaedaDadhich3}). It can be also developed
in the framework of the Lovelock theory of the general type
\cite{Lovelock}; we plan to do this comparing with known results
(see. e.g., \cite{AFrancavigliaR,KatzLivshits}).

\subsection*{Acknowledgments} The author is very grateful to Joseph
Katz for the prolonged and fruitful discussions, due to which many
points have been clarified and the presentation has been
significantly improved. The author expresses also his gratitude to
Alikram Aliev for deep explanations of his works, very useful
conversations and recommendations for future development of the
results. The work is supported by the grant No. 09-02-01315-a of the
Russian Foundation for Basic Research.

\section{\bf Erratum, submitted to CQG}

\bigskip

\noindent Petrov A N 2009 Three types of superpotentials for
perturbations in the Einstein-Gauss-Bonnet gravity {\em Class.
Quantum Grav.} {\bf 26} 135010 (16pp) (Paper the above.)

In this short communication we want to correct a mistake in our the
above work. For the coefficient $m$ in (2.9) it was taken a wrong
order of a covariant differentiation of $g$ in the first case (first
line). The right writing is
  \bea
  \hat m_\sig{}^{\alf\tau} & \equiv & -\frac{1}{2\k}\l\{
 \l[{{\di \lag_c} \over {\di (\Bar D_\alf g^a)}} -
 \Bar D_\beta \l({{\di \lag_c} \over {\di (\Bar D_\alf \Bar D_\beta g^a)}}\r)\r]
 \l.g^a\r|^\tau_\sig \r.\nonumber \\ & - & \l.
 {{\di \lag_c} \over {\di (\Bar D_\tau \Bar D_\alf g^a)}}
\Bar D_\sig g^a +
 {{\di \lag_c} \over {\di (\Bar D_\beta \Bar D_\alf g^a)}}
 \Bar D_\beta (\l.g^a\r|^\tau_\sig)\r\}\, .
\m{(+4+)}
 \eea
In the result, some formulae of the canonical approach have to be
changed. Thus instead of the formulae (3.2) and (3.3), which define
the general canonical superpotential in EGB gravity, one must to
write, respectively,
  \bea
\hat \imath^{\alf\beta}_C  &= & {}_{E}\hat \imath^{\alf\beta}_C
+{}_{GB}\hat \imath^{\alf\beta}_C
 \nonumber\\ &=&
{1\over \k}\big({\hat g^{\rho[\alf}\Bar D_{\rho} \xi^{\beta]}} +
\hat g^{\rho[\alf}\Delta^{\beta]}_{\rho\sig}\xi^\sig\big)
 \nonumber\\ &- & \frac{2\alf\sqrt{-g}}{\k}
 \l\{\Delta^{\rho}_{\lam\sig}R_\rho{}^{\lam\alf\beta} +
4\Delta^{\rho}_{\lam\sig}g^{\lam[\alf}R^{\beta]}_\rho +
\Delta^{[\alf}_{\rho\sig}g^{\beta]\rho}R\r\}\xi^\sig
 \nonumber\\&-&
\frac{2\alf\sqrt{-g}}{\k}\l\{R_\sig{}^{\lam\alf\beta} +
 4 g^{\lam[\alf}R^{\beta]}_\sig + \delta_\sig^{[\alf}g^{\beta]\lam}R
 \r\}\BD_\lam \xi^\sig\, ,
 \m{CanSupEGB}
 \eea
 \bea
\Bar{\hat \imath^{\alf\beta}_C}  &= & {}_{E}\Bar{\hat
\imath^{\alf\beta}_C} +{}_{GB}\Bar{\hat \imath^{\alf\beta}_C}
 \nonumber\\ &=&
{1\over \k} \Bar {D^{[\alf} \hat\xi^{\beta]}}
 -
\frac{2\alf}{\k}\l\{\Bar{\hat R}_\sig{}^{\lam\alf\beta} +
 4 \Bar{g^{\lam[\alf}\hat R^{\beta]}_\sig} + \delta_\sig^{[\alf}
 \Bar{g^{\beta]\lam}\hat R}
 \r\}\BD_\lam \xi^\sig\, .
 \m{BarCanSupEGB}
 \eea
One can see that the mistake influences only on the GB part. Next,
instead of the formulae (4.3), (4.4) and (4.5) one has the following
formulae, respectively,
 \bea
 {\hat\imath}^{01}_C &+& \k^{-1}\lam^{[0}\hat d^{1]}_{DKO}
 = \frac{\sqrt{-\Bar g}}{2\k r}\l[\frac{r\Bar{f}'}{2}\l(\frac{f}{\Bar f}+\frac{\Bar
f}{f} \r) - (f-\Bar f)(D-2)\r]\nonumber \\&+&
 \frac{\alf\sqrt{-\Bar g}}{\k r^2}(D-2)(f-\Bar f)\l(f'- rf''\r) \nonumber \\
& -&\frac{\alf\sqrt{-\Bar g}}{\k r^3}(D-2)(D-3)(f-1)
\l[\frac{r\Bar{f}'}{2}\l(\frac{f}{\Bar f}+\frac{\Bar f}{f} \r)
-(f-\Bar f)(D-2)\r],
  \m{Can-i+d}
 \eea
 \bea
 {\hat\imath}^{01}_C &+& \k^{-1}\lam^{[0}\hat d^{1]}_{KL}
 = \frac{\sqrt{-\Bar g}}{2\k r}\l[\frac{r\Bar{f}'}{2}\l(\frac{f}{\Bar f}+\frac{\Bar
f}{f} \r) - (f-\Bar f)(D-2)\r]\nonumber \\&+&
 \frac{\alf\sqrt{-\Bar g}}{\k r^2}(D-2)(D-3)(f-\Bar f)f'
 \nonumber \\
& -&\frac{\alf\sqrt{-\Bar g}}{\k r^3}(D-2)(D-3)(f-1)
\l[\frac{r\Bar{f}'}{2}\l(\frac{f}{\Bar f}+\frac{\Bar f}{f} \r)
-(f-\Bar f)(D-4)  \r],
 \m{Can-i+dKL}
 \eea
 \be
\Bar{\hat\imath^{01}_C}  = \frac{\sqrt{-\Bar g}}{2\k }\Bar{f}'
 - \frac{\alf\sqrt{-\Bar g}}{\k r^2}(D-2)(D-3)\Bar{f}'(\Bar f-1)
\,.
  \m{BarCan-i+d}
 \ee
The above three expressions define the canonical superpotential for
the static spherically symmetric solutions of the Schwarzschild-like
form.

Turn to the formula (5.6), which is used for calculating the mass of
the Schwarzschild-anti-de Sitter black hole. It has a place for all
the three approaches examined in the paper and is the main paint in
carrying out the applications. It turns out that this asymptotic
formula is conserved for the {\em here corrected} canonical
superpotential in the EGB gravity. Inversely, it is surprising that
(5.6) has a place for the previous wrong canonical superpotential!
Also, the mistake has no an influence on writing the Belinfante
superpotential because it does not depend on $m$-coefficient; and
the mistake has no an influence on writing the symmetrical
superpotential because it depends on the coefficient $N$ in (2.21),
where the order of particular derivatives of $g$ is not important.
At last, we note that, in spite of the mistake, all the discussions
and conclusions of the paper are right.

To conclude the comment we remark the following. How we understand
now, after analyzing the mistake, there exist {\em different
possibilities} to choose an order of covariant derivatives of $g$ in
definitions of coefficients, like $m$ in (2.9) and $n$ in (2.10). In
the other words, the choice in $n$ defines the choice in $m$. Only,
in all the cases $n$ and $m$ have to satisfy the standard system of
the N{\oe}ther identities. In future, we plan to study this
possibility constructing a new set of conserved quantities
(including new set of superpotentials) by a non-contradictive way.

\ed